\documentclass[a4paper,12pt]{article}
\usepackage{amsmath}
\usepackage{graphicx}

\begin{document}

\begin{center}

{\Large \bf Muon $(g-2)$ from the bulk neutrino field in a warped extra dimensional model}\\[20mm]

R. S. Hundi\footnote{E-mail address: tprsh@iacs.res.in},
Sourov Roy\footnote{E-mail address: tpsr@iacs.res.in} and
Soumitra SenGupta\footnote{E-mail address: tpssg@iacs.res.in}\\
Department of Theoretical Physics,\\
Indian Association for the Cultivation of Science,\\
2A $\&$ 2B Raja S.C. Mullick Road,\\
Kolkata - 700 032, India.\\[20mm]

\end{center}

\begin{abstract}

In the Randall-Sundrum model, a bulk neutrino field in the 5-dimensional
space-time can give rise to tiny Dirac masses to neutrinos. In such a scenario, we have
computed the contribution of the bulk neutrino field to the anomalous magnetic moment
$(g-2)_\mu$ of muon. We have computed this contribution in the 't Hooft-Feynman gauge and
have found that the contribution has the right sign to fit the current discrepancy between
the experiment and the standard model value of $(g-2)_\mu$. We have also studied possible
constraints on the model parameters by including contributions to $(g-2)_\mu$ from other
sources such as bulk gravitons.

\end{abstract}

\newpage

\section{Introduction}

The models in extra dimensions have been proposed for solving the hierarchy between
the electroweak and Planck scales \cite{ADD,RS}. Among these the Randall-Sundrum (RS)
model assumes the existence of two 3-branes connected by one extra spatial
dimension and the metric in this model is non-factorizable \cite{RS}. The extra spatial coordinate has been
orbifolded by the symmetry $S^1/Z_2$ and one of the 3-branes can be identified as the
visible brane and the other as the Planck brane. The warp factor in this model suppresses
any Planck scale quantities into electroweak scale on the visible brane. In the
RS model, all the standard model fields are assumed to be confined on the visible brane and
only gravity propagates in the bulk of the five dimensions. Subsequently, the RS model was
generalized to include other bulk fields in order to explain physical quantities such as
neutrino masses and mixing pattern which normally cannot be explained within the standard
model of elementary particles.

The deficit in the solar and atmospheric neutrino flux has given evidence for non-zero masses to
neutrinos \cite{nuexp}. Fitting to the data of solar and atmospheric neutrino experiments, the following 
set of mass-square differences are obtained: $\Delta m_{\rm solar}^2 = m_2^2-m_1^2 \approx 7.6 \times 
10^{-5}$ eV$^2$ and $\Delta m_{\rm atm}^2 = |m_3^2 - m_{1,2}^2| \approx 2.5\times 10^{-3}$ eV$^2$ 
\cite{global_fit}. Here, $m_{1,2,3}$ are the mass eigenvalues of the 3 active neutrinos. 
Apart from the mass-square differences an upper limit on the neutrino masses have been found through other
experiments. Tritium $\beta$-decay puts an upper limit on the neutrino mass scale to be of the order of
2 eV \cite{b-decay}. Whereas from the cosmological observations, the sum of the three neutrino masses needs 
to be less than about 1 eV \cite{cosmos}. Since all the above experiments suggest tiny values for neutrino masses, 
perhaps a different mechanism should be operational for neutrino mass generation as compared to other fermion 
masses.

To explain the smallness of neutrino masses in the framework of RS model, an additional singlet
neutrino field has been proposed, which like gravity field, propagates in the entire bulk of
space-time \cite{Gross_Neub}. The wave function of the bulk neutrino field is extended in the extra spatial 
dimension.  The boundary conditions for this wave function can be chosen in such a way that it will
have a very small overlap on our visible brane, resulting in tiny masses for neutrinos. The phenomenology
of this model is determined through the Kaluza-Klein (KK) modes of the bulk neutrino field. Since the
bulk neutrino field is singlet under the standard model gauge group, detection of KK modes of this
field is challenging in the collider experiments. However, the loop effects due to these fields
to any physically observable quantity can give us some hints about its existence. Here, we study
one of such observable quantities, namely the anomalous magnetic moment of muon. 

At the tree level, the anomalous magnetic moment of muon, $g$-factor, has a value of 2 and
radiative contributions give some corrections to it. Hence, it can be quantified as
$a_\mu = (g-2)_\mu/2$. For a review on $(g-2)_\mu$, see \cite{Zhang,Jeger_Nyff}.
The world average value of $(g-2)_\mu$ after the experiment E821
at the Brookhaven National Laboratory is as given below \cite{g-2exp}
\begin{equation}
a_\mu^{\rm EXP} = 11659208.0(6.3)\times 10^{-10},
\end{equation}
which is obtained with a precision of 0.54 parts per million.
Various groups have computed the theoretical value
for $(g-2)_\mu$ in the standard model. Most of the groups have found a discrepancy between
the experiment and the corresponding standard model value of $(g-2)_\mu$ at about $3\sigma$
level \cite{g-2res}. Here, we take this difference as follows \cite{Jeger_Nyff}
\begin{equation}
\Delta a_\mu = a_\mu^{\rm EXP} - a_\mu^{\rm SM} = (29\pm 9)\times 10^{-10}.
\label{E:expg-2}
\end{equation}
The above difference would indicate existence of new physics. Moreover, since this difference
is positive, the contribution due to new physics to $(g-2)_\mu$ should yield a net positive value.

As explained before, in the model of Ref. \cite{Gross_Neub}, KK neutrinos can give some contribution
to $(g-2)_\mu$, which we have computed in this work. Previously, some work in this direction has been
done in Ref. \cite{Mc_Ng}, where the authors have obtained a negative contribution to the
$(g-2)_\mu$ by adopting mass insertion approximation in the unitary gauge.
In the present work, by carrying out an exact analysis in the mass eigenstate basis, we have computed the
contribution from the bulk neutrino field to the $(g-2)_\mu$ in the 't Hooft-Feynman
gauge and scan the parameter space in the region consistent with experimental bounds. 
We have found that the bulk neutrino contribution
to $(g-2)_\mu$ has the right sign to fit the above mentioned discrepancy in $(g-2)_\mu$.
We have also incorporated contributions
from other sources such as gravitons to the $(g-2)_\mu$. Finally,
we have studied the constraints that may arise from these various sources of
$(g-2)_\mu$ in the parameter space defined in \cite{Gross_Neub}. Recently, in \cite{Iyer_Vemp},
constraints have been obtained from lepton flavor violation by considering a set of models
in the RS frame work.

The paper is organized as follows. In the next section, we give a brief description
of the warped extra dimensional model for neutrino masses \cite{Gross_Neub}. In Sec. 3,
we have computed the contribution of bulk neutrino field to $(g-2)_\mu$ in the 't Hooft-Feynman
gauge. In this section, we have also performed a detailed phenomenological study of our obtained
expression for $(g-2)_\mu$. In Sec. 4, we describe possible constraints on the model parameters 
of \cite{Gross_Neub} by including contributions to $(g-2)_\mu$ from other sources such as gravitons. 
We conclude in Sec. 5.

\section{Dirac neutrinos in the warped extra dimensional model}

As described before, the model \cite{Gross_Neub} is based on the RS model, where an additional
singlet neutrino field is introduced in the bulk of the 5-dimensional space-time. The metric in this
model is
\begin{equation}
ds^2 = e^{-2\sigma(\phi)}\eta_{\mu\nu}dx^\mu dx^\nu + r_c^2 d\phi^2,
\end{equation}
where $\sigma(\phi)=kr_c|\phi|$,
$r_c$ is the compactification radius of the fifth dimension and $k$ is an energy scale
of the order of Planck scale, $M_P$. Here, $\phi$ is an angular coordinate representing the fifth
dimension. Due to the $S^1/Z_2$ orbifold symmetry, $\phi$ varies from 0 to $\pi$ and the 3-branes
located at these points are called Planck and visible branes, respectively. In this model,
the invariant action for a singlet bulk fermion $\Psi$ is as given below.
\begin{equation}
S = \int d^4x\int d\phi \sqrt{G}\left\{E^A_a\left[\frac{i}{2}\left(\bar{\Psi}\gamma^a\partial_A\Psi
-\partial_A\bar{\Psi}\gamma^a\Psi\right)+\frac{\omega_{bcA}}{8}\bar{\Psi}\{\gamma^a,\sigma^{bc}\}\Psi\right]
-M_b{\rm sgn}(\phi)\bar{\Psi}\Psi\right\},
\end{equation}
where $G$ is the determinant of 5-dimensional metric $G_{AB}$, $E^A_a$ is the inverse vierbein
and $\omega_{bcA}$ is the spin connection. The small case letters, $a,b,c$, run over flat 5-dimensions
and upper case Roman letters run over curved 5-dimensional space-time. Here, $M_b$ is a bulk mass parameter
which is ${\cal O}(M_P)$. The above action is invariant
under $\phi\to -\phi$, which should follow due to the $Z_2$ orbifolding of the RS model. This symmetry
is known as $\phi$-parity which sets some boundary conditions on the wave functions
of $\Psi$. The bulk field $\Psi$ decomposes into KK modes in the 4-dimensional world, which can
be written as
\begin{equation}
\Psi_{L,R}(x,\phi)=\sum_n\psi_n^{L,R}(x)\frac{e^{2\sigma}}{\sqrt{r_c}}f_n^{L,R}(\phi),
\label{E:KKmode}
\end{equation}
where $\Psi_{L,R}=\frac{1}{2}(1\mp\gamma_5)\Psi$, $\psi^{L,R}_n(x)$ are the left- and
right-handed KK modes in the 4-dimensions and $f^{L,R}_n(\phi)$ are its corresponding wave functions.
In order to have the following canonically normalized action for the KK modes in 4-dimensions
\begin{equation}
S_{\rm eff} = \sum_n\int d^4x\left\{\bar{\psi}_n(x)i\gamma^\mu\partial_\mu\psi_n(x)
-m_n\bar{\psi}_n(x)\psi_n(x)\right\},
\end{equation}
the wave functions $f_n^{L,R}$ should satisfy the following conditions
\begin{equation}
\int_\epsilon^1 dtf_m^{L*}(t)f_n^L(t) = \int_\epsilon^1 dtf_m^{R*}(t)f_n^R(t) = \delta_{mn},
\quad
(\pm t\partial_t - \nu)f_n^{L,R}(t) = -x_ntf_n^{R,L}(t).
\label{E:coneqs}
\end{equation}
Here, $m_n$ are the masses of KK modes.
In the above equations, the following change of variables have been done : $t=\epsilon e^{\sigma(\phi)}$,
$\epsilon = e^{-kr_c\pi}$ and $f_n^{L,R}(\phi)\to\sqrt{kr_c\epsilon}f_n^{L,R}(t)$. This change
of variables imply that $t=1$ corresponds to the visible brane. The unknown
quantities in the above equation are :
\begin{equation}
\nu = \frac{M_b}{k},\quad x_n = \frac{m_n}{k}e^{kr_c\pi}.
\end{equation}
The $\phi$-parity, which is described above,
imposes the following boundary conditions on the wave functions : $f_m^{L*}(\epsilon)f_n^R(\epsilon)
=f_m^{L*}(1)f_n^R(1)=0$. The coupled differential equations of Eq. (\ref{E:coneqs}) can be exactly solved and for
$n>0$, the wave functions can be expressed in the form of Bessel functions. The wave function
for the zeroth mode can be written as
\begin{equation}
f_0^{L,R}(t)=f_0^{L,R}(1)t^{\pm\nu},\quad |f_0^{L,R}(1)|^2=\frac{1\pm 2\nu}{1-\epsilon^{1\pm 2\nu}}.
\end{equation}
The zeroth mode wave function of the right-handed field is highly suppressed on the
visible brane, for $\nu>\frac{1}{2}$. Because of this reason, by choosing the boundary condition
as $f_n^L(\epsilon)=f_n^L(1)=0$, we can achieve a very small overlap of the singlet bulk field
on the visible brane, for $\nu>\frac{1}{2}$. Using this mechanism, in the next paragraph, we
describe the neutrino masses in this model.

The invariant action for neutrino Yukawa interaction is
\begin{equation}
S_Y = -\int d^4x\sqrt{-g_{\rm vis}}\{\hat{Y}_5\bar{L}_0(x)\tilde{H}_0(x)\Psi_R(x,\pi) + {\rm h.c.}\},
\end{equation}
where $L_0$ is a left-handed lepton doublet and $\tilde{H}_0$ is the conjugate of the Higgs doublet.
Here $\hat{Y}_5$ has mass dimensions of $\frac{1}{\sqrt{M_P}}$ and the metric on the
visible brane is
$\left(g_{\rm vis}\right)_{\mu\nu} = e^{-2\sigma(\pi)}\eta_{\mu\nu}$ and
$g_{\rm vis} = {\rm det}\left(\left(g_{\rm vis}\right)_{\mu\nu}\right)$.
Both the lepton and Higgs doublets need to have the following rescaling :
$L_0 = e^{\frac{3}{2}\sigma(\pi)}L$, $H_0 = e^{\sigma(\pi)}H$, in order to have
canonical kinetic terms for these fields.
After substituting the KK mode expansion, eq. (\ref{E:KKmode}), the invariant action becomes
\begin{equation}
S_Y = -\int d^4x\sum y_n \bar{L}\tilde{H}\Psi^R_n(x) + {\rm h.c.},
\quad y_n = \sqrt{k}\hat{Y}_5f_n^R(1) = Y_5f_n^R(1).
\label{E:SY}
\end{equation}
Here, $Y_5$ is an order one parameter. In the basis, $\Psi_L^\nu = (\nu_L,\Psi^L_1,
\cdots,\Psi_n^L)$ and $\Psi_R^\nu = (\Psi_0^R,\Psi^R_1,\cdots,\Psi^R_n)$, we have
\begin{equation}
S_Y = -\int d^4x \overline{\Psi_L^\nu}M\Psi^\nu_R + {\rm h.c.},
\quad
M = \left(\begin{array}{cccc}
                   vy_0 & vy_1 & \cdots & vy_n \\
                   0    & m_1  & \cdots & 0    \\
                   0    & 0    & \ddots & 0    \\
                   0    & 0    & 0      & m_n
                   \end{array}\right).
\label{E:mixmas}
\end{equation}
Here, the vacuum expectation value of the Higgs field is $v=$ 174 GeV. Now, define
\begin{equation}
\Psi_L^\nu = U^L\nu^{\rm kk}_L, \quad \Psi_R^\nu = V^R\nu^{\rm kk}_R,
\label{E:KK}
\end{equation}
then the physical masses are given by
\begin{equation}
(U^L)^\dagger M V^R = M_{\rm diag}.
\end{equation}
While diagonalizing the matrix $M$, the masses $m_n$ of KK modes are determined
by the equation $J_{\nu-\frac{1}{2}}(x_n)=0$. However, the masses $m_n$ are not the
physical masses of KK modes, since they get correction due to mixing with the left-handed
lepton doublet after the electroweak symmetry breaking. After diagonalizing the matrix $M$,
the lowest mass eigenvalue can be shown to be proportional to $y_0$ which yields a
very small value due to small overlap of the wave function $f_0^R(1)$.
The mass of light neutrino can be shown to be \cite{Gross_Neub}
\begin{equation}
m_\nu = \sqrt{2\nu -1}Y_5\epsilon^{(\nu-\frac{1}{2})}v.
\label{E:mnu}
\end{equation}
So far we have assumed the presence of only one bulk neutrino field and it generates one light
Dirac neutrino mass. As explained in the previous section, to fit the neutrino oscillation data,
we need at least two non-zero neutrino mass eigenstates. Hence, in the model of Ref. \cite{Gross_Neub}
we have to propose more than one bulk neutrino field. However, it has been argued in \cite{Gross_Neub}
that to cancel the anomalies related to $\phi$-parity, only even number of singlet bulk fields can
be introduced into the model. So, by introducing two bulk neutrino fields with slightly different bulk
mass parameters, $M_b$, we can fit both the solar and atmospheric neutrino mass scales. In this
picture the third neutrino has exactly zero mass. However, by introducing 4 bulk neutrino fields in
an analogous way, we can generate all the three light neutrino masses.

\section{Contribution of KK neutrinos to $(g-2)_\mu$}

We have shown in the previous section that to generate neutrino masses, the KK modes
of bulk neutrino field have some mixing with the left-handed lepton doublet. This mixing
would lead to gauge as well as Yukawa type interactions of muon with the physical KK modes
of neutrinos. As a result of this, the KK modes of neutrinos give some contribution to
$(g-2)_\mu$, which we will describe shortly. It is to be noticed here that we have computed
the contribution of bulk neutrino to $(g-2)_\mu$ in the 't Hooft-Feynman gauge, hence the
interactions of muon with Nambu-Goldstone states are necessary. For the sake of completeness,
below we describe interaction terms involving the physical KK modes of bulk neutrino
field. The interaction terms of other fields are unchanged from that of
the standard model.\\
\underline{\bf The coupling of $\mu W^+\nu_i^{\rm kk}$} :\\
In this model, the interaction term of $W$-boson with charged lepton and neutrino
would come from the 5-dimensional invariant action of the kinetic energy of the lepton doublet.
For our particular case of muon field, these interactions are as given below.
\begin{equation}
S_K \ni \int d^4x \frac{g}{\sqrt{2}}[\overline{\nu_{\mu L}}\gamma^\mu\mu_LW^+_\mu
+\overline{\mu_L}\gamma^\mu\nu_{\mu L}W^-_\mu].
\end{equation}
After using Eq. (\ref{E:KK}) we get
\begin{equation}
S_K \ni \int d^4x \frac{g}{\sqrt{2}}\sum_i\left[\left(U^L_{1i}\right)^*\overline{\nu^{\rm kk}_i}\gamma^\mu
\frac{1-\gamma_5}{2}\mu W^+_\mu + U^L_{1i}\overline{\mu}\gamma^\mu\frac{1-\gamma_5}{2}\nu^{\rm kk}_iW^-_\mu\right].
\end{equation}
\underline{\bf The coupling of $\mu G^+\nu^{\rm kk}_j$} :\\
Without loss of generality, we can go to a basis where charged lepton Yukawa couplings
are diagonalized. In this basis, the necessary Lagrangian for our purpose is
\begin{equation}
{\cal L} = -Y_\mu\bar{L}_\mu H\mu_R- \sum_jy_j\bar{L}_\mu\tilde{H}\Psi^R_j + {\rm h.c.}.
\end{equation}
Here $L_\mu$ is the left-handed lepton doublet containing muon. The second term in the above equation
is from Eq. (\ref{E:SY}). We can use
$Y_\mu = m_\mu/v$ where $m_\mu$ is the mass of muon. After using Eq. (\ref{E:KK}), we can show that
\begin{eqnarray}
{\cal L} &\ni & -\frac{m_\mu}{v}\sum_j\overline{\nu^{\rm kk}_j}G^+\left(U^L_{1j}\right)^*\frac{1+\gamma_5}{2}\mu
+\sum_{i,j}\overline{\nu^{\rm kk}_j}G^+y_i^*\left(V^R_{i,j}\right)^*\frac{1-\gamma_5}{2}\mu
\nonumber \\
&& -\frac{m_\mu}{v}\sum_j\overline{\mu}G^-U^L_{1j}\frac{1-\gamma_5}{2}\nu^{\rm kk}_j
+\sum_{i,j}\overline{\mu}G^-y_iV^R_{ij}\frac{1+\gamma_5}{2}\nu^{\rm kk}_j.
\end{eqnarray}

The interaction terms, as described above, generate some contribution to $(g-2)_\mu$ at
one loop level, which are shown in Fig. 1.
\begin{figure}[!h]
\begin{center}

\includegraphics[]{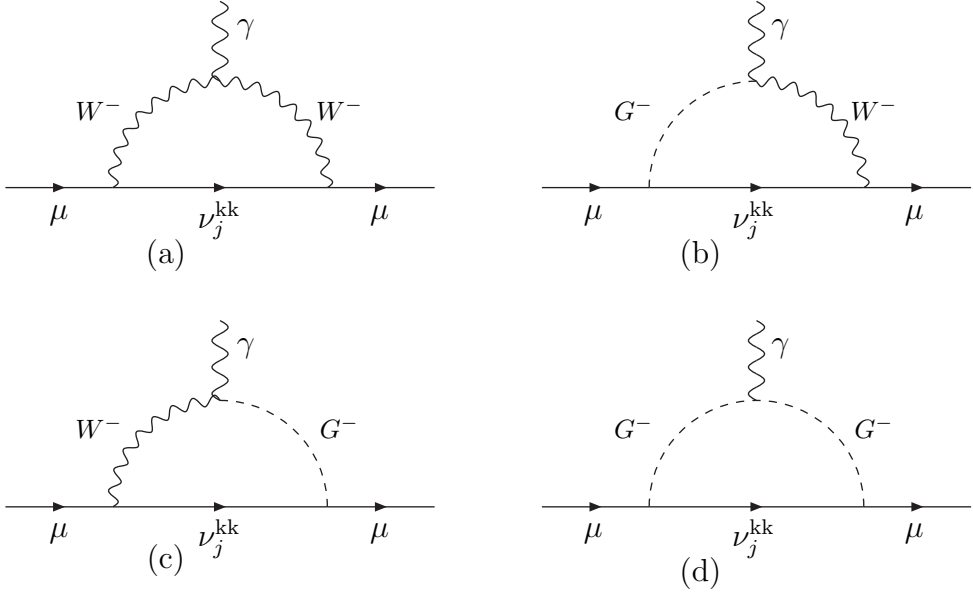}

\end{center}
\caption{The contribution of KK neutrinos to $(g-2)_\mu$. The momentum convention is that
the outgoing muon has $p^\prime$, the incoming muon has $p$ and the momentum on the photon
line is $q=p^\prime - p$, which is incoming.}
\end{figure}
The amplitude of any of the Fig. 1 is of the form $i{\cal M} = -ie\bar{u}(p^\prime)\Gamma^\mu u(p)\epsilon_\mu$,
where $\epsilon_\mu$ is the polarization of photon and $e$ is the positron charge. After using the Gordon identity,
$\bar{u}(p^\prime)\gamma^\mu u(p) = \bar{u}(p^\prime)\left[\frac{(p+p^\prime)^\mu}{2m_\mu}+
\frac{i\sigma^{\mu\nu}q_\nu}{2m_\mu}\right]u(p)$, where $q=p^\prime - p$, 
we can put the interesting part of the amplitude as
\begin{equation}
\Gamma^\mu = \gamma^\mu F_1(q^2)+\frac{i\sigma^{\mu\nu}q_\nu}{2m_\mu}F_2(q^2) +\cdots .
\end{equation}
The contribution to $(g-2)_\mu$ is $\Delta a_\mu = \frac{(g-2)_\mu}{2} = F_2(0)$.
%

The contribution from Fig. 1(a) to $(g-2)_\mu$ is
\begin{eqnarray}
\Delta a^{(1)}_\mu &=& \frac{m_\mu^2}{32\pi^2}\sum_{j=2} g^2U^L_{1j}\left(U^L_{1j}\right)^*\frac{1}{(m^{\rm kk}_j)^2}
F_a(x_j), \quad x_j=\frac{m_W^2}{(m^{\rm kk}_j)^2},
\nonumber \\
F_a(x) &=& \frac{1}{(1-x)^4}\left[-\frac{31}{6}+\frac{19}{2}x-\frac{11}{2}x^2+\frac{7}{6}x^3-(3-x)\ln(x)\right].
\end{eqnarray}
%
The sum of the contributions from Figs. 1(b) and 1(c) to $(g-2)_\mu$ is as follows
\begin{eqnarray}
\Delta a^{(2)}_\mu &=& \frac{m_\mu^2}{32\pi^2}\sum_{j=2} g^2U^L_{1j}\left(U^L_{1j}\right)^*\frac{1}{(m^{\rm kk}_j)^2}
F_b(x_j), \quad x_j=\frac{m_W^2}{(m^{\rm kk}_j)^2},
\nonumber \\
F_b(x) &=& \frac{1}{(1-x)^3}\left[-\frac{3}{2}+2x-\frac{1}{2}x^2-\ln(x)\right].
\end{eqnarray}
%
The contribution to $(g-2)_\mu$ from Fig. 1(d) is
\begin{eqnarray}
\Delta a^{(3)}_\mu &=& \frac{m_\mu^2}{16\pi^2}\sum_{i=2,j=2,k=2}
\left\{-\left[\frac{m_\mu^2}{v^2}U^L_{1j}\left(U^L_{1j}\right)^*+y_iy^*_k
V^R_{ij}\left(V^R_{kj}\right)^*\right]\frac{1}{(m^{\rm kk}_j)^2}F_c(x_j)
\right. \nonumber \\
&& \left. +\frac{1}{vm^{\rm kk}_j}\left[U^L_{1j}y^*_k\left(V^R_{kj}\right)^*+\left(U^L_{1j}\right)^*y_kV^R_{kj}\right]
F_d(x_j)\right\}, \quad x_j=\frac{m_W^2}{(m^{\rm kk}_j)^2},
\nonumber \\
F_c(x) &=& \frac{1}{(1-x)^4}\left[\frac{1}{3}+\frac{1}{2}x-x^2+\frac{1}{6}x^3+x\ln(x)\right],
\nonumber \\
F_d(x) &=& \frac{1}{(1-x)^3}\left[\frac{1}{2}-\frac{1}{2}x^2+x\ln(x)\right].
\end{eqnarray}
It is to be noticed that in the contribution to $(g-2)_\mu$ from all the plots of Fig. 1,
the summation in indices over $i,j,k$ is from 2, since we have to
subtract the light neutrino contribution which exists in the standard model.

Here we comment on our results on the contribution of bulk neutrino to $(g-2)_\mu$. As already
explained before that we have computed the
contribution from Fig. 1 in the 't Hooft-Feynman gauge. The first three
diagrams of Fig. 1 have given positive contributions, while the Fig. 1(d) has given both
positive as well as negative contributions. In our numerical analysis, which we
present below, we have found that in most of the parameter space the contribution from
Fig. 1(d) is dominant and it gives a net positive contribution. Specifically, we have seen that
the negative and positive contributions are comparable to each other but the magnitude of the former
one is at least an ${\cal O}(1)$ less than the later one in $\Delta a^{(3)}_\mu$.
Another comment is that, since there are two insertions of Yukawa couplings in Fig. 1(d) we
would expect the contribution to $(g-2)_\mu$ to increase with $Y_5$. In our numerical
analysis we have found that this is true but there are some exceptions to this, which
we will explain in our numerical results.

The total contribution of one bulk neutrino field to the $(g-2)_\mu$ is
\begin{equation}
\Delta a^{N}_\mu = \Delta a^{(1)}_\mu + \Delta a^{(2)}_\mu + \Delta a^{(3)}_\mu
\end{equation}
The above contribution
is mainly dependent on $\nu$, $kr_c$ and the 5-dimensional bulk mass parameter $M_b$,
which altogether determine the KK masses of the
bulk neutrino field. Apart from this, $\Delta a^N_\mu$ also depends on the dimensionless
parameter $Y_5$ which determine the elements of the unitary matrices $U^L$ and $V^R$.
On the other hand, the light neutrino mass eigenvalue due to one bulk neutrino field,
depends on $\nu$, $kr_c$ and $Y_5$. As described in Sec. 1, from the neutrino oscillation data we have
some idea on the magnitude of neutrino masses. For instance, in the hierarchical
pattern of neutrinos at least two mass eigenvalues should be :
$m_{\rm solar}=\sqrt{\Delta m^2_{\rm solar}}\approx 0.01$ eV and $m_{\rm atm}=\sqrt{\Delta m^2_{\rm atm}}\approx 0.05$ eV.
In this case, the third
neutrino can have either zero mass or its mass should be less than of the order of $m_{\rm solar}$.
In the case of degenerate neutrinos, all the three neutrinos should have a nearly equal mass and
the common mass eigenvalue should be sufficiently larger than $m_{\rm atm}$. Since an upper
bound from the cosmological observations indicate that the sum of the three neutrino masses should
be less than about 1 eV \cite{cosmos}, we take the common mass to be as $m_{\rm deg}\approx 0.3$ eV.
By fixing the mass eigenvalues of light neutrinos, we may eliminate $\nu$ as independent
variable. The parameter $Y_5$ should be ${\cal O}(1)$ and here we take its value in the range
0.1 to 2.5. Here the lower limit of 0.1 is due to the naturalness argument and the upper limit
is due to perturbativity constraints. In order not to hit the Landau pole, the Yukawa couplings
$y_n$ should be less than $\sqrt{4\pi}$. For $n>0$, $|f_n^R(1)| = \sqrt{2}$ and hence $Y_5$
should be less than $\sqrt{2\pi}\approx$ 2.5. As for the bulk mass parameter $M_b$, we take
its value close to the Planck scale. We take the values of $kr_c$ to be around 12, since for this
set of values we can produce TeV scale masses on the visible brane from the exponential warping.
Finally, in our numerical analysis, we have allowed a 50$\times$50 mixing mass matrix of
Eq. (\ref{E:mixmas}). After diagonalizing this matrix we get 49 KK neutrino mass modes which we
have summed in the $\Delta a^N_\mu$. We have checked that the above mentioned number of KK modes
in $\Delta a^N_\mu$ is sufficient and its value vary insignificantly by further increasing
the number of KK modes.

In Fig. 2 we have plotted the contribution due to one bulk neutrino field to $(g-2)_\mu$.
\begin{figure}[!h]
\begin{center}

\includegraphics[height=2.5in, width=2.5in]{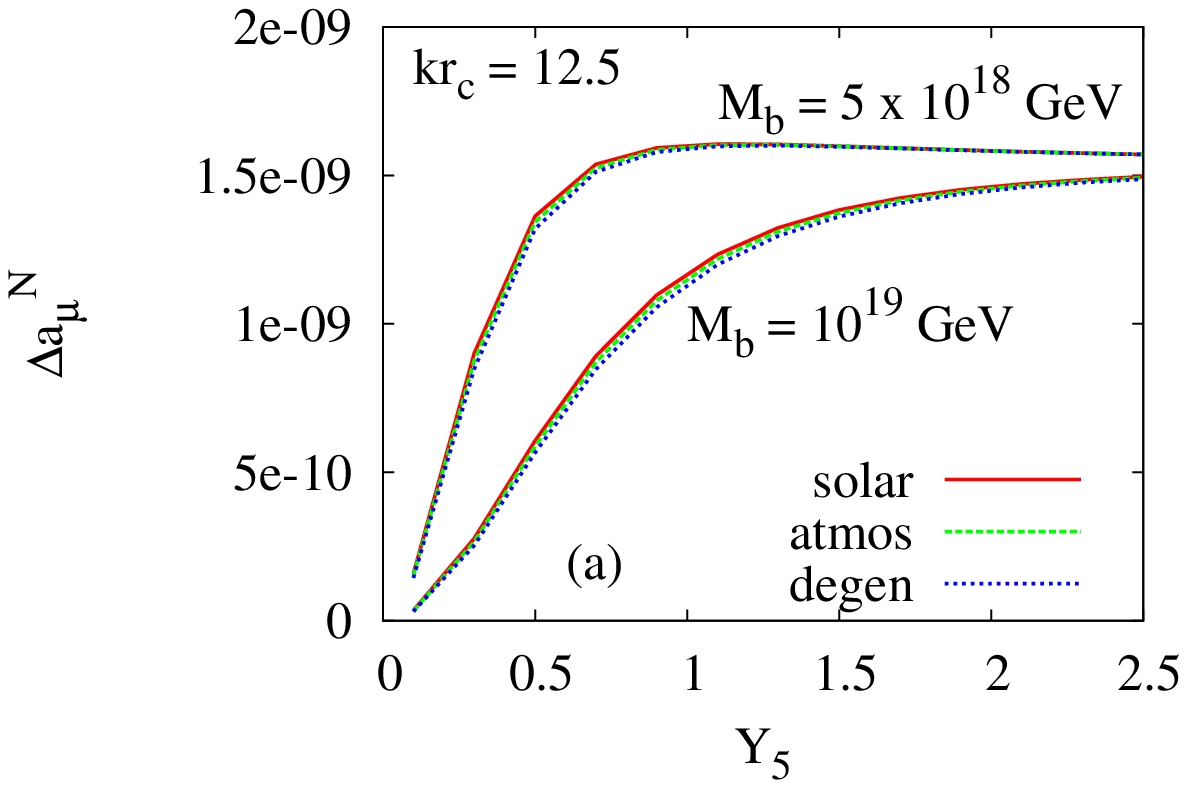}
\includegraphics[height=2.5in, width=2.5in]{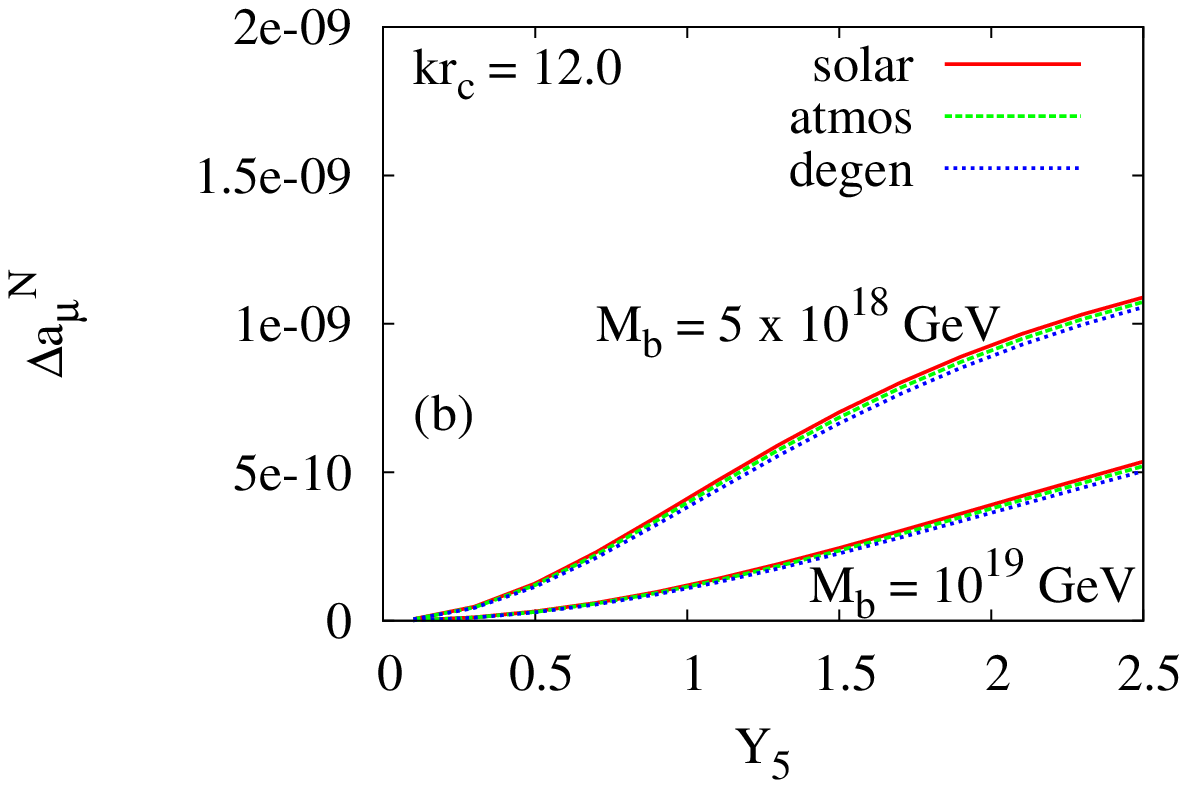}

\end{center}
\caption{$\Delta a^N_\mu$ has been plotted as a variable of $Y_5$. In these plots
the three curves which are closely stacked together represents whether the light
neutrino mass fits the solar, or atmospheric, or degenerate mass. The values of $kr_c$
and $M_b$ have been varied in both these plots.}
\end{figure}
In Fig. 2(a) we have fixed $kr_c$ = 12.5. In this plot, the upper three curves are due to
a bulk mass of $M_b = 5\times 10^{18}$ GeV and the remaining three lower curves are for $M_b = 10^{19}$ GeV.
The three curves are arising depending on whether the light neutrino mass eigenvalue
fits the solar, or atmospheric, or degenerate mass eigenvalue. The value of $\nu$ is
not shown in these plots, since as explained before it is not an independent parameter. The meaning
of curves in Fig. 2(b) are same as that of Fig. 2(a), except that $kr_c$ has taken a
value 12.0 in Fig. 2(b). For one particular value of $Y_5$, the contribution due to
a bulk neutrino field which fits the solar, or atmospheric, or degenerate neutrino is nearly the same.
This is evident in both the Figs. 2(a) and 2(b), where the three curves due to different neutrino
mass eigenvalues are close to each other.
We can understand the reason for this as follows. For fixed values of
$Y_5$, $kr_c$ and $M_b$, the change in the value of $\Delta a_\mu^N$ from one curve to the other
curve can happen only
due to the change in the neutrino mass eigenvalue. The ratio of neutrino mass eigenvalues
between degenerate and solar cases could be at most by a factor of ${\cal O}(10)$. From Eq. (\ref{E:mnu}),
it can be seen that the neutrino mass eigenvalue is related to $\nu$ dominantly through
exponential factor,
so we need only a change of $\sim$0.1 in $\nu$ to produce a change of factor ${\cal O}(10)$
in the mass eigenvalue. These exact numerical values in the change of $\nu$ due to change in the neutrino
mass eigenvalue can be seen in Tab. 1, where ranges of $\nu$ are given for extreme
values of $Y_5$. Such a small change in $\nu$ produces a slight change in the KK masses of neutrino
fields, and hence a slight change to the $\Delta a^N_\mu$.
However, a closer examination would reveal that the
amount of $\Delta a_\mu^N$ where solar mass is fitted is slightly more than that for atmospheric
mass which is even slightly more than that for degenerate mass. The change in these three cases
could be at most in the first decimal place of $\Delta a_\mu^N$ value. The values of $\nu$ in these three cases
would be different even though $Y_5$ could be same. We have presented the ranges of $\nu$ for
the cases of $kr_c$ = 12.5 and 12.0 in Tab. 1.
\begin{table}[!h]
\begin{center}
\begin{tabular}{||c|c|c||}\hline
 & $kr_c = 12.5$ & $kr_c = 12.0$\\\hline
$m_{\rm sol}$ & $\nu\sim (1.22 - 1.31)$ & $\nu\sim (1.25 - 1.34)$ \\
$m_{\rm atm}$ & $\nu\sim (1.18 - 1.26)$ & $\nu\sim (1.21 - 1.30)$ \\
$m_{\rm deg}$ & $\nu\sim (1.13 - 1.22)$ & $\nu\sim (1.16 - 1.25)$ \\\hline
\end{tabular}
\end{center}
\caption{Ranges of $\nu$ for different values of $kr_c$ and for different
masses of light neutrino mass eigenvalues. In a particular range of $\nu$, the
left- and right-end corresponds to $Y_5 = 0.1$ and $Y_5 = 2.5$, respectively.}
\end{table}
The ranges of $\nu$ in Tab. 1 are for extreme values of $Y_5$ of
0.1 and 2.5. Since the mass eigenvalue of neutrino, eq. (\ref{E:mnu}), does not depend
on the bulk mass parameter $M_b$, the range of $\nu$ is also independent of this parameter.
From the plots of Fig. 2, we have realized that there is some sensitivity
between the light neutrino mass eigenvalue and the corresponding $\Delta a^N_\mu$
value. However, for the same value of $Y_5$, this sensitivity is so small
that we may not determine the neutrino mass eigenvalue based on the
$\Delta a^N_\mu$ value.

Next, we present results on how the $\Delta a_\mu$ varies by changing the $kr_c$ as
well as the bulk mass parameter $M_b$, which is given in Fig. 3.
\begin{figure}[!h]
\begin{center}

\includegraphics[height=2.5in, width=2.5in]{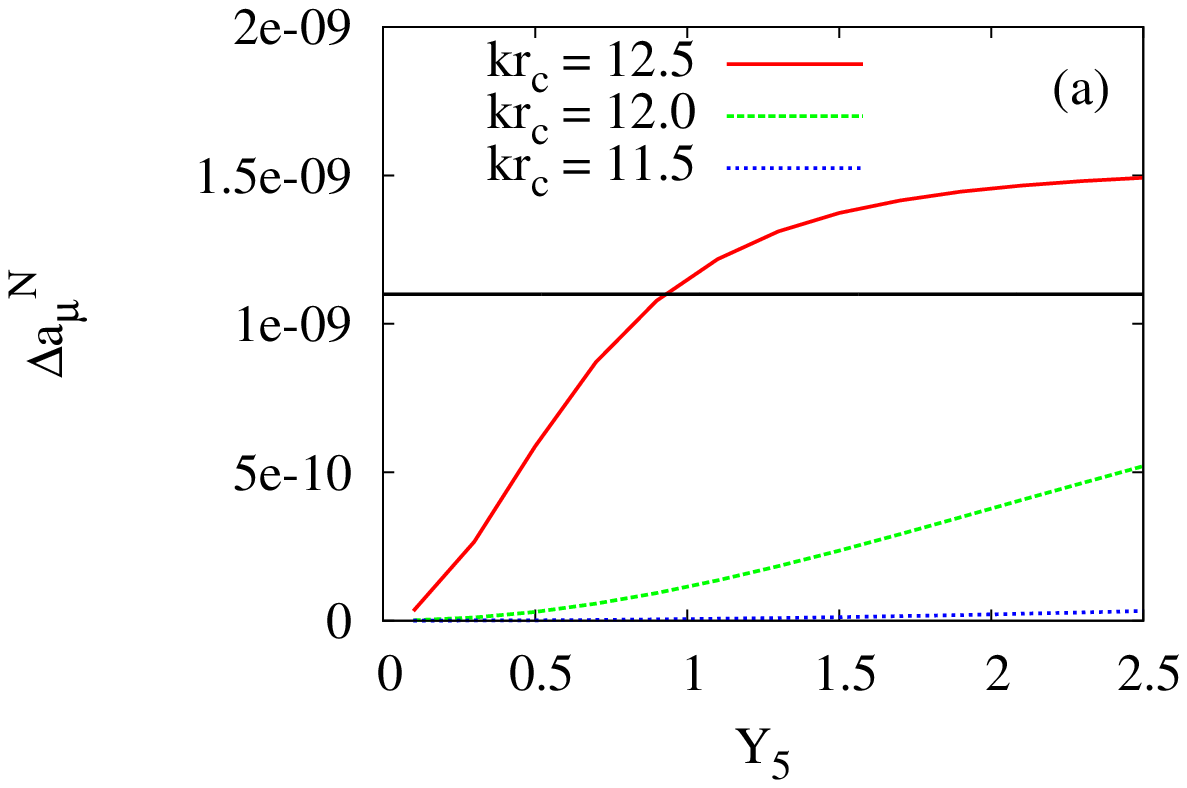}
\includegraphics[height=2.5in, width=2.5in]{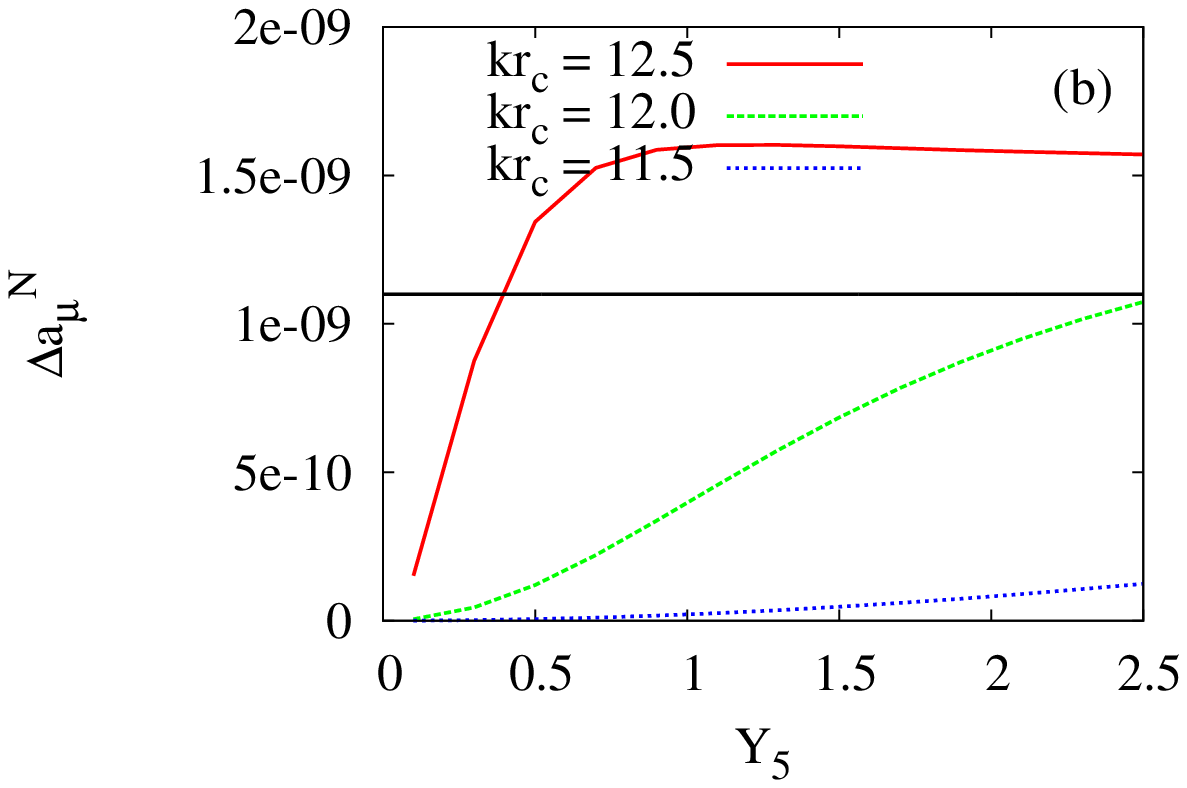}

\includegraphics[height=2.5in, width=2.5in]{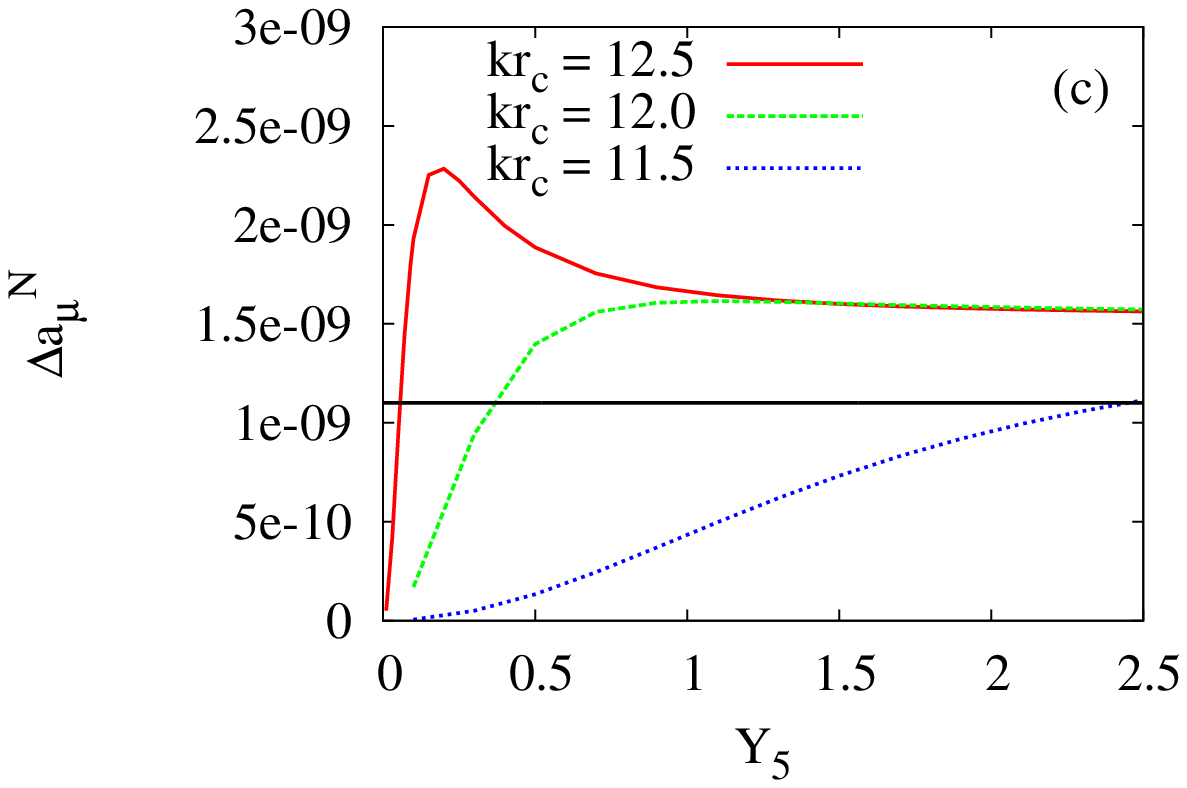}

\end{center}
\caption{$\Delta a^N_\mu$ has been plotted against $Y_5$ for different values
of $kr_c$. In the top-left and top-right plots, the value of $M_b$ has been taken as
$10^{19}$ GeV and $5\times 10^{18}$ GeV, respectively. In the lower-middle
plot, $M_b=10^{18}$ GeV. The horizontal line in all these plots indicates the
$2\sigma$ lower limit of $\Delta a_\mu$, see the text for details.}
\end{figure}
In the plots of Fig. 3, we have fixed the light neutrino mass eigenvalue to
be the atmospheric scale $m_{\rm atm}$. The choice of neutrino mass scale do not
make much difference in numerical values which we have argued around Fig. 2. From the plots of Fig. 3
we can understand that
by increasing the value of $kr_c$ the exponential warping would decrease the
KK masses of neutrinos and hence the contribution to $\Delta a^N_\mu$ would
increase. It can also be noticed from Fig. 3 that
the lower the bulk mass parameter the larger the contribution is to the $\Delta a^N_\mu$,
which is evident since the masses of KK neutrinos would become lower. For low values
of $Y_5$, the contribution to $\Delta a^N_\mu$ is low and it is increasing and in some cases
it may be saturated for large enough $Y_5$.
We have noticed that all the curves would be saturated for some large enough $Y_5$. For example, 
the curves for $kr_c=12.0$ in Figs. 3(a) and 3(b) would saturate for $Y_5$ around 50 and 20, respectively.
The dependence of $\Delta a_\mu^N$ on $Y_5$ is
somewhat complicated as can be noticed from the theoretical expression given above. However,
we can notice that Fig. 1(d) has two insertions of Yukawa couplings and hence we can predict
that $\Delta a^{(3)}_\mu$ should increase with $Y_5$. On the other hand, $Y_5$ also affect the
contributions of the first three plots of Fig. 1 by determining the elements of $U^L$ and $V^R$
and also the physical KK masses of neutrinos.
Numerically we have seen that for large value of $Y_5$, the contribution
from Fig. 1(d) is always 2 orders greater than that due to other plots of Fig. 1. The contribution
from Fig. 1(d) is given in the form of $\Delta a^{(3)}_\mu$, from which we can see that there
is a partial cancellation due to positive and negative contributions of $\Delta a^{(3)}_\mu$.
We have numerically seen that $\Delta a^{(3)}_\mu$ goes to a saturation value for large enough $Y_5$.
Since in the plots of Fig. 3, we have fixed the neutrino mass eigenvalue to atmospheric scale, and
the mass eigenvalue has an exponential dependence on $\nu$, we have found that after large enough $Y_5$
the change in $\nu$ would be far less compared to the change in $Y_5$. Hence the values of $m_n$ which
determine the KK masses of neutrinos would almost be saturated.
The elements of $U^L$ and $V^R$ would change with $Y_5$, however, numerically we have seen
that the net sum of the various KK modes is saturated after large enough value of $Y_5$.
In the plot of Fig. 3(c),
for the case of $kr_c = 12.5$ and $M_b= 10^{18}$ GeV the amount of $\Delta a^N_\mu$ would peak
at around $Y_5\sim 0.1$. In this particular case the contribution from $\Delta a^{(1)}_\mu$
is significantly dominant at around $Y_5\sim 0.1$. However, in this case the lowest KK mass of neutrino
is around 30 GeV. In other cases where the lowest KK mass is at least few 100 GeV,
$\Delta a^{(3)}_\mu$ would give the dominant contribution. We have noticed these facts
purely from numerical values.

The results described above show that the contribution from a single bulk
neutrino field in five dimensions can easily fit, depending on the values of parameters,
the $2\sigma$ deviation of $\Delta a_\mu$. From Eq. (\ref{E:expg-2}) we take the
$2\sigma$ deviation as
\begin{equation}
\Delta a_\mu = (1.1 - 4.7)\times 10^{-9}.
\label{E:2sig}
\end{equation}
However, to be specific, the results of Fig. 3 indicate that the contribution
to $\Delta a_\mu$ is most likely towards the lower end of the above $2\sigma$
deviation or even less than this depending on the values of $Y_5$.
In the case of $kr_c = 11.5$, the contribution to $\Delta a_\mu$ is much below the
$2\sigma$ limit for $M_b =$ $10^{19}$ GeV or $5\times 10^{18}$ GeV. But keeping
$M_b=10^{18}$ GeV and for $Y_5\sim 2$, we get a value of $\Delta a_\mu\sim 10^{-9}$.
Since this is only from one bulk neutrino field and in a realistic scenario we
need at least two bulk neutrino fields, so
we can fit the above $2\sigma$ deviation by adjusting the $M_b$ and $Y_5$,
even for a low value of $kr_c = 11.5$.

In the above analysis we have presented our results due to the existence of one bulk neutrino
field. By introducing a second bulk neutrino field, the additional main parameters that the
second field would carry are its 5-dimensional bulk mass parameter, $M_b$, and its Yukawa coupling
$Y_5$ to the muon. We can convince ourselves that the expression for $\Delta a_\mu$ due to this second 
field would be same as that of the first field, but replace the above said parameters accordingly. Now, 
consider a realistic scheme where there are two bulk neutrino fields which fits both the solar and
atmospheric neutrino mass scales. In order to fit these neutrino mass scales, we may choose
their Yukawa couplings to be nearly same but the values of $\nu$ would be slightly different
for these fields, which can be understood from Tab. 1. This would result in slight difference
in the respective values of their bulk parameters $M_b$. Hence the total contribution to $(g-2)_\mu$
due to these two bulk fields would almost be the twice of the contribution from a single bulk field.

\section{Constraints on the model parameters by including graviton contribution}

In the previous section, we have shown that the contribution due to bulk neutrino
field can fit the $2\sigma$ deviation of $\Delta a_\mu$. The fits in the previous section
would in fact set limits on the parameter space of the model in \cite{Gross_Neub}, if this is the only
source for $(g-2)_\mu$. However, the contribution
due to the bulk neutrino may get further constraints due to the presence of other
sources in the 5-dimensional warped model, such as from graviton or radion fields.
Since gravity exists in the whole space-time, we cannot ignore its contribution to
the $\Delta a_\mu$ \cite{Kim_Song}. Similarly, the length of the fifth dimension
should be dynamically generated and this leads to the presence of radion field.
We have found that the contribution from gravitons to $\Delta a_\mu$ is significantly
larger than that of the radion contribution \cite{Das_Mahanta}. Below, we show
how the contribution from the gravitons to $\Delta a_\mu$ would lead to some
constraints on the model parameters of \cite{Gross_Neub}.

Regarding the graviton contribution to $(g-2)_\mu$, it has been computed in the case
of RS model in \cite{Hooman_etal,Kim_Song}. However, the expression given in \cite{Hooman_etal}
has been estimated by assuming all the standard model fields
in the bulk of five dimensions. Since in the model of \cite{Gross_Neub}, only the singlet bulk
neutrino and gravity are allowed in the bulk of space-time, we confine to the results given
in \cite{Kim_Song}, where it is claimed to be done for the case of the original RS model.
In \cite{Kim_Song}, an expression for the contribution of KK gravitons to the $(g-2)_\mu$
at one loop level is given, which is as follows
\begin{equation}
\Delta a_\mu^{g} = \frac{5}{16\pi^2}\left(\frac{m_\mu}{\Lambda_\pi}\right)^2n_c,
\end{equation}
where $n_c$ is the number of KK gravitons and $\Lambda_\pi = e^{-kr_c\pi}M_P$. It has
also been argued in \cite{Kim_Song} that the number of KK gravitons are bounded by some
unitarity constraints due to $\gamma\gamma$ elastic scattering, where it has been shown to
be $n_c\leq 10 - 100$. By taking the Planck scale as $M_P = 10^{19}$ GeV, the contribution
due to a single KK graviton are as follows : $(kr_c,\Delta a_\mu^{g})=(12.2,6.8\times 10^{-9}),
(12.1,3.64\times 10^{-9}),
(12.0,1.94\times 10^{-9}), (11.7, 2.95\times 10^{-10}), (11.3, 2.39\times 10^{-11})$. Comparing
these values with the $2\sigma$ range of Eq. (\ref{E:2sig}), anything above the $kr_c = 12.1$ can
be ruled out purely from the KK graviton contribution to $(g-2)_\mu$. For $kr_c =12.0$, only
two KK gravitons can exists. Whereas, for $kr_c =$ 11.7 and 11.3, of the order of 10 and 100 number
of KK gravitons can exist, respectively. Although it may be inappropriate to assume the
existence of only one or two KK gravitons in the universe, nevertheless, by the unitarity bounds
of \cite{Kim_Song} it seems to be consistent. In these cases we show how this assumption
would put bounds on the bulk neutrino parameters of \cite{Gross_Neub}.

For $kr_c=12.1$, there can exist only one KK graviton to fit the $2\sigma$ deviation of
$\Delta a_\mu$. In this case, an
amount of $\Delta a_\mu^{\rm max}-\Delta a_\mu^{g} =$ (4.7 - 3.64)$\times 10^{-9}$ = 1.06$\times 10^{-9}$
can be shared by contributions due to other sources. It has been shown in \cite{Das_Mahanta} that
for radion mass greater than about 200 GeV, its contribution to $\Delta a_\mu$ would be 
less than $10^{-9}$. For simplicity, we ignore this contribution by assuming radion mass to
be greater than 200 GeV. Then in the degenerate masses of light neutrinos, since at least
4 bulk neutrinos should exist, the contribution from any single bulk neutrino should be
$\Delta a_\mu^{N}\leq 2.65\times 10^{-10}$. On the other hand, if the neutrino mass
pattern is hierarchical then at least two bulk neutrinos should exist. By assuming that
each bulk neutrino contributes by the same amount to $(g-2)_\mu$, we can then put a bound
$\Delta a_\mu^{N}\leq 5.3\times 10^{-10}$. An upper bound on $\Delta a_\mu^{N}$ would
imply an upper limit on $Y_5$ and thereby an upper bound on the corresponding
$\nu$ of the bulk neutrino. In Tab. 2 we show conservative bounds on these parameters
for various values of bulk parameters $M_b$.
\begin{table}
\begin{center}
\begin{tabular}{||c||c|c||c|c|c|c||} \hline
& \multicolumn{2}{|c||}{Degenerate neutrinos}
& \multicolumn{4}{|c||}{Hierarchical neutrinos}\\\hline
$M_b$ & $Y_5$ & $\nu$ & $Y_5^a$ & $\nu^a$ & $Y_5^s$ & $\nu^s$ \\ \hline
$10^{19}$ GeV & 1.2 & 1.22 & 1.9 & 1.28 & 1.8 & 1.32 \\
$5\times 10^{18}$ GeV & 0.6 & 1.2 & 0.9 & 1.26 & 0.9 & 1.31 \\
$10^{18}$ GeV & 0.1 & 1.15 & 0.14 & 1.21 & 0.14 & 1.25 \\ \hline
\end{tabular}
\end{center}
\caption{Upper limits on $Y_5$ and $\nu$ for different values of $M_b$ and in different
patterns of neutrino masses. In the case of hierarchical neutrinos, the suffix $a$ and $s$
represent for atmospheric and solar neutrinos, respectively.}
\end{table}
In the case of $kr_c=12.0$, two KK gravitons can exist to fit the $(g-2)_\mu$.
The above analysis done for $kr_c=12.1$ can be repeated for $kr_c=12.0$ and
we can get bounds on the parameters $Y_5$ and $\nu$ accordingly. However, if we include
the contribution of the order of 10 KK gravitons then $kr_c$ should be close to 11.7. In this
case, by comparing the curves for $kr_c=11.5$ of Fig. 3, for $M_b$ close to $10^{18}$ GeV,
a single bulk neutrino can contribute an order of $10^{-9}$ to $(g-2)_\mu$. So for this
choice of $M_b$ we can get upper limits on $Y_5$ and $\nu$. However, if the number of
KK gravitons is of the order of 100 then $kr_c$ should be around 11.3. For this low value
of $kr_c$, a single bulk neutrino contribution would be at most $6\times 10^{-10}$ even
for $M_b=10^{18}$ GeV. Hence the constraints on the bulk neutrino parameters would be less
stringent in this case. Finally, to comment on the radion contribution to $(g-2)_\mu$,
the constraints on the bulk neutrinos which are described above may become even severe
if the radion mass is less than about 200 GeV.

In the previous paragraph we have described constraints on the bulk neutrino parameters
which are arising purely from the $2\sigma$ deviation of $\Delta a_\mu$. However, these constraints
can be even stringent by including other observable quantities such as $BR(\mu\to e\gamma)$.
The current experiments have not found the decay $\mu\to e\gamma$ and put an upper limit
on its branching ratio as $BR(\mu\to e\gamma)<2.4\times 10^{-12}$ at 90$\%$ CL \cite{mutoegamma}.
In the context
of the warped model \cite{Gross_Neub}, this decay channel has been studied in \cite{Kitano}.
To satisfy the upper bound on $BR(\mu\to e\gamma)$ and to get an appreciable contribution
from $\Delta a_\mu^N$, we may have to fine-tune the Yukawa couplings.
It is interesting to study these effects but this is out of the scope of the current work.
Finally, we comment that in this work we have studied correlation between neutrino mass eigenstate
and $(g-2)_\mu$ for a single generation. We hope to extend  this to include all the generations in a future
work. Some related work in the context of different models has been studied in \cite{othg-2wk}.

\section{Conclusions}

The RS model, which is based on the warp geometry, is elegant to explain the
hierarchy between the Planck and electroweak scales \cite{RS}. Subsequently it was shown that
such model can offer possible explanation of the origin of small neutrino masses when the right
handed component of the neutrino is allowed to propagate in the bulk \cite{Gross_Neub}.
The signals of such a model lies in the detection of KK modes of the bulk neutrino field. In this
work, we have studied an indirect signal of these KK modes, {\it i.e.} the virtual effects of these
KK modes to the anomalous magnetic moment of muon, $(g-2)_\mu$. We have computed the contribution
of bulk neutrino to $(g-2)_\mu$ at one loop level in the 't Hooft-Feynman gauge. Comparing our result
with the experimental limits, we have found that this contribution comes with the right sign so as
to fit with experimental data. Moreover, in some of the parameter space of the model in 
\cite{Gross_Neub}, the $2\sigma$ discrepancy of $(g-2)_\mu$
can be accommodated. We have also studied how the bulk neutrino contribution would be restricted by
including the graviton contribution to $(g-2)_\mu$.

\section*{Acknowledgment}
SR acknowledges the hospitality provided by Prof. Alejandro Ibarra and the Cluster of Excellence 
for Fundamental Physics ``Origin and Structure of the Universe" at the Technical University in Munich (TUM) 
during the final stages of this work.

\end{document}